\documentclass[journal]{IEEEtran}

\usepackage{amssymb}
\usepackage{amsmath}
\usepackage{graphicx}
\usepackage{setspace}
\usepackage{color}
\usepackage{balance}
\usepackage{flushend}

\begin{document}
%
\title{Application of Expurgated PPM to Indoor Visible Light Communications - Part II: Access Networks}

\author{{Mohammad Noshad,~\IEEEmembership{Student Member,~IEEE}, and Ma\"{\i}t\'{e} Brandt-Pearce,~\IEEEmembership{Senior Member,~IEEE}}
\thanks{This paper was presented in part at IEEE Globecom 2013.

 Mohammad Noshad (mn2ne@virginia.edu) and Ma\"{\i}t\'{e} Brandt-Pearce (mb-p@virginia.edu) are with Charles L. Brown Department of Electrical and Computer Engineering, University of Virginia, Charlottesville, VA 22904.}
}

\markboth{}{Shell \MakeLowercase{\textit{et al.}}: Bare Demo of IEEEtran.cls for Journals}

\maketitle

\begin{abstract}
Providing network access for multiple users in a visible light communication (VLC) system that utilizes white light emitting diodes (LED) as sources requires new networking techniques adapted to the lighting features. In this paper we introduce two multiple access techniques using expurgated PPM (EPPM) 
that can be implemented using LEDs and support lighting features such as dimming. Multilevel symbols are used to provide $M$-ary signaling for multiple users using multilevel EPPM (MEPPM). 
Using these multiple-access schemes we are able to control the optical peak to average power ratio (PAPR) in the system, and hereby control the dimming level. In the first technique, the $M$-ary data of each user is first encoded using an optical orthogonal code (OOC) assigned to the user, and the result is fed into a EPPM encoder to generate a multilevel signal. The second multiple access method uses sub-sets of the EPPM constellation to apply MEPPM to the data of each user. While the first approach has a larger Hamming distance between the symbols of each user, the latter can provide higher bit-rates for users in VLC systems using bandwidth-limited LEDs.
\end{abstract}


\section{Introduction}
\PARstart{V}{isible} light communications (VLC) is an appealing technology for network access in indoor environments since it is immune to radio-frequency (RF) interference, has low impact on human health, and is able to provide a high data-rate connection \cite{Indoor-OWC-Haas11}. It has been proposed as an alternative to WiFi to provide high-speed access for tablets, phones, laptops and other devices in indoor spaces such as offices, homes, airplanes, hospitals and convention centers.
The application of VLC to indoor networking not only requires the capability to provide simultaneous connection for a large number of users, but also to meet the requirements of the lighting system, mainly dimming.
In this paper we introduce two techniques to provide high-speed multiple access for simultaneous users in a VLC system.

Integrating VLC networks with illumination systems imposes limitations on the modulations and networking techniques that can be used. White light emitting diodes (LEDs) are the most common optical sources that are used in VLC systems, and modulation schemes that can be used with these devices are limited. Because of the structure of these LEDs and their inherent nonlinearity, implementing modulation and multiple-access approaches that require frequency-domain processing is expensive and complicated. Therefore, time-based modulations, specially pulsed techniques, are the preferred modulation technique in LED-based VLC systems. Dimming is an important feature of indoor lighting systems through which the illumination level can be controlled. Including dimming in VLC system requires further constraints on the multiple-access schemes that can be used.
A practical VLC network should support various optical peak to average power ratios (PAPR) so that, for a fixed peak power, the average power, which is proportional to the illumination, can be regulated.

Optical code division multiple access (OCDMA) is a networking technique that provides multiple access by assigning binary signature patterns to users \cite{OCDMA}. Among various OCDMA forms that have been proposed, direct-sequence OCDMA is of most interest for indoor VLC system, since it can be implemented by simply turning the LEDs on and off \cite{OWC-OCDMA12}. In this type of OCDMA, binary sequences with special cross-correlation constraints, such as optical orthogonal codes (OOC), are used to encode the data of users in the time-domain \cite{OCDMA1-89}. Codewords of an OOC are binary sequences that meet a given correlation constraint \cite{OOC-89}. The application of OOCs to VLC networks requires codes with a wide range of parameters for different dimming levels, which may not be practical for a network with a large number of users.

In part I of this two-part paper \cite{VLC-JLT-I-13}, we propose to apply expurgated pulse-position modulation (EPPM) \cite{EPPM12} and multilevel-EPPM (MEPPM) \cite{Multilevel-EPPM12}, both based on balanced incomplete block designs (BIBD), to indoor VLC systems in which the dimming can be done by simply changing the generating BIBD code.
%
%
%
In this part II of the paper, two networking methods are proposed based on MEPPM to provide multiple access for simultaneous users in a VLC system. These two techniques, which can be considered  synchronous OCDMA methods, enable users in a VLC network to have high-speed access to the network. In the first method we assign one OOC codeword to each user in order to encode its $M$-ary data. For each user, every bit of this encoded binary sequence is multiplied by a BIBD codeword, and then the OOC-encoded BIBD codewords are added to generate a multilevel signal. Hence, the PAPR of the transmitted data can be controlled by changing the code-length to code-weight ratio of the BIBD code. In the second technique, a subset of BIBD codewords is assigned to each user, and then the MEPPM scheme is used to generate multi-level symbols using the assigned codewords. In this approach, users can have different bit-rates by partitioning the BIBD code into unequal-size subsets.

The organization of the rest of the paper is as follows. In Section II, we describe the indoor VLC network. The two networking methods to provide simultaneous access for multiple users are proposed in Section III. The performance of the proposed techniques are compared using numerical results in Section IV. Finally, Section V concludes the paper.

\section{System Description}

\begin{figure} [!t]
    \begin{center}
    {\includegraphics[width=3.2in]{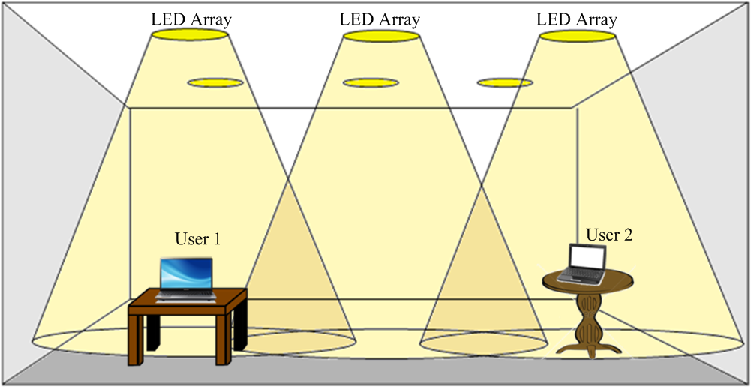}}
    \end{center}
    \vspace*{0.0 in}
    \caption{Configuration of a VLC network with LED arrays and two users.}
    \label{VLC}
    \vspace*{-0.0 in}
    \end{figure}

This section describes the principles of a VLC network.
In LED-based VLC systems, both lighting and communications needs can be addressed at the same time. The downlink configuration of a VLC network is shown in Fig.~\ref{VLC}, where arrays of white LEDs are used as sources.
Here, we consider LED arrays as access points, and we assume that all LED arrays in a room are synchronous and transmit the same data.
For each user, usually the strongest received power is the one that corresponds to the signal received from the direct path of the closest LED array, and therefore, it is considered as the main signal and the main data source. In situations where the direct path to the main source is blocked, the data can be retrieved using the multipath signals received from non-line-of-sight (NLOS) paths \cite{VLC-JLT-I-13}.
In this paper, VLC is assumed to be used only for the downlink, and another independent system, such RF or infrared (IR) communications, is used for the lower data-rate uplink channel to avoid self-interference from the full-duplex communication. 

\begin{figure} [!t]
    \vspace*{-0.0 in}
    \begin{center}
    {\includegraphics[width=3.1in]{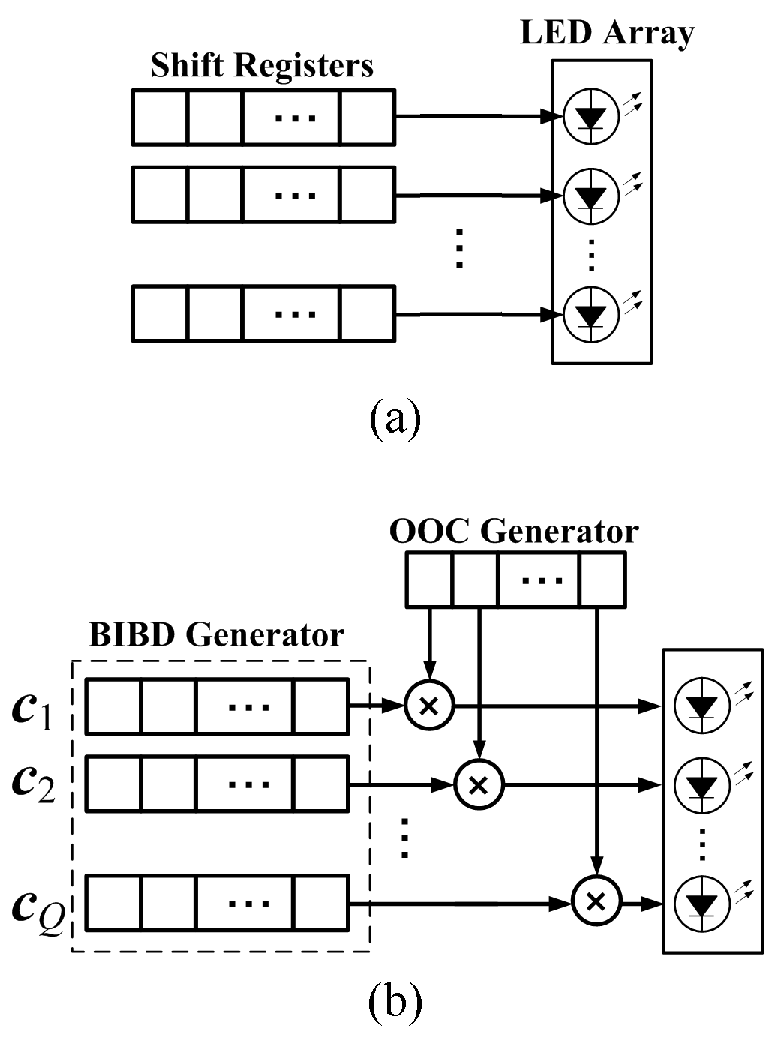}}
    \end{center}
    \vspace*{-0.0 in}
    \caption{Transmitter structure and symbol generation using shift registers for (a) MEPPM, and (b) coded-MEPPM.}
    \label{C-MEPPM Transmitter}
    \end{figure}

In \cite{Multilevel-EPPM12}, multilevel signals are constructed by combining multiple BIBD codes, which are then used as modulation symbols in multilevel EPPM (MEPPM). Each symbol-time is divided into $Q$ equal time-slots, and then subsets of LEDs within the array are turned on and off according to a BIBD codeword.
In this work, BIBDs are used to generate the multiple-access codewords. We utilize a BIBD code with parameters ($Q$,$K$,$\lambda$), where $Q$, $K$ and $\lambda$ denote the code-length, the Hamming weight and the cross-correlation between any two codewords, respectively. The $m$th codeword is the binary vector $\mathbf{c}_m=(c_{m1}, c_{m2},\dots, c_{mQ} ) =\mathbf{c}_1^{(m)}$, $m = 1, 2, \dots, Q $, where the notation $\mathbf{x}^{(m)}$ is the $m$th cyclic shift of the vector $\mathbf{x}$. The following relation holds between $\mathbf{c}_{m}$'s and is referred to as the fixed cross-correlation property \cite{Noshad11}:
    \begin{align}\label{Cross Correlation Property}
        \sum_{i=1}^{Q} c_{mi} c_{ni} = \left\{ \begin{array}{lll}
                                        K            &; \, m = n, \\
                                        \lambda      &; \, m \neq n
                                        \end{array}
              \right.  .
    \end{align}
    %
Four different MEPPM schemes are introduced in \cite{Multilevel-EPPM12}, two of which (called type-I and type-II MEPPM) have a PAPR of $Q/K$ and are used in this paper. An implementation using on-off modulated LEDs is shown in Figure~\ref{C-MEPPM Transmitter}(a).

\begin{figure} [!t]
    \begin{center}
    {\includegraphics[width=3.2in]{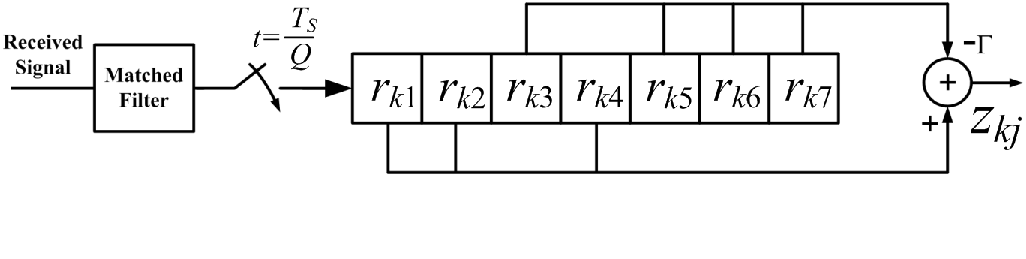}}
    \end{center}
    \vspace*{0.0 in}
    \caption{Receiver for MEPPM using a (7,3,1)-BIBD code.} 
    \label{EPPM Decoder}
    \vspace*{-0.0 in}
    \end{figure}

The structure of a simple receiver for single user EPPM and MEPPM systems using a shift-register is shown in Fig.~\ref{EPPM Decoder}. In this receiver, for symbol-epoch $k$, the sampled data at the output of a pulse-matched filter, $\mathbf{r_k}$, is stored in a shift register, and then is circulated inside it to generate vector $\mathbf{z}_k=(z_{k1},z_{k2},\dots,z_{kQ})$, $z_{kj}=\langle  \mathbf{r}_k ,\mathbf{c}_j \rangle$, at the output of the differential circuit, where $\langle \mathbf{x}, \mathbf{y} \rangle$ denotes the dot product of the vectors $\mathbf{x}$ and $\mathbf{y}$.
In this figure, $T_s$ is the symbol time and $\Gamma=\lambda/(w-\lambda)$. The wires of the lower branch are matched to the first codeword of the BIBD code, $\mathbf{c}_1$, and those of the upper branches are matched to its complement. This receiver is equivalent to the correlation decoder, which is shown to be the optimum decoder for additive-white-Gaussian-noise (AWGN) channels in \cite{EPPM12} and for shot-noise channels in \cite{VLC-JLT-I-13}.


\section{Multiple Access for Indoor Optical Networks}
Indoor optical networks must be able to provide simultaneous access for multiple users.
%
Optical code division multiple access (OCDMA) can be used to fulfill this need.
In VLC systems, since the access points are illumination sources and are the same for all users, synchronous OCDMA techniques can be used for the downlink. In \cite{OWC-OCDMA12}, synchronous time-spreading OCDMA using optical orthogonal codes (OOC) and on-off keying (OOK) was studied. Because of the limited bandwidth of LEDs and the long length of OOCs needed, the data-rate for each user is low. An efficient technique to increase the data-rate in OOC-based OCDMA systems is $M$-ary modulation using cyclic shifts of the OOC codewords, which is called code cycle modulation (CCM) in \cite{OCDMA-M-ary-06}. In this modulation, any cyclic shift of an OOC codeword with length $L$ is considered as a symbol, and therefore, the bit-rate is increased by a factor of $\log_2 L$. This technique is highly susceptible to synchronization errors, as is any technique using cyclic codes.

A limitation of using OOCs in VLC systems is their incompatibility with the dimming feature. For a CCM OCDMA system that uses an OOC with length $L$ and weight $w$, the PAPR is $L/w$. Furthermore, for an OOC code with length $L$, weight $w$, and cross-correlation $\alpha$, the number of codewords, which must be at least as large as the number of users, $N$, is bounded by the Johnson bound \cite{OOC-89}
    \begin{align}\label{Johnson Bound}
        N \leq \Bigg\lfloor \frac{1}{w} \bigg\lfloor \frac{L-1}{w-1} \Big\lfloor \frac{L-2}{w-2} \dots \Big\lfloor \frac{L-\alpha}{w-\alpha} \Big\rfloor \dots \Big\rfloor  \bigg\rfloor \Bigg\rfloor.
    \end{align}
Changing the pulse duty-cycle is not possible for bandwidth-limited sources, while changing the pulse amplitude requires a complex tuner circuit due to the source nonlinearity.
Therefore, in an OCDMA network with a given number of users, changing the PAPR requires employing a new OOC code, which, considering (\ref{Johnson Bound}), may not be possible for low PAPRs. As an example, for a PAPR of 2, we should have $L=2w$, which implies that $\alpha > w^2/L = w/2$ and corresponds to systems with large $N$'s. Codes with these parameters are not only difficult and highly complicated to design, but also have high interference collision probability and poor performance. Furthermore, to change either the number of user or the dimming level it is necessary to change the OOC code, which requires a large database of OOC codes.

In this section, two networking methods based on MEPPM are introduced in order to not only provide multiple access for different users in an indoor VLC network, but also provide $M$-ary transmission for each user so that a higher data-rate can be achieved.

    \begin{figure} [!t]
    \vspace*{-0.0 in}
    \begin{center}
    {\includegraphics[width=3.0in]{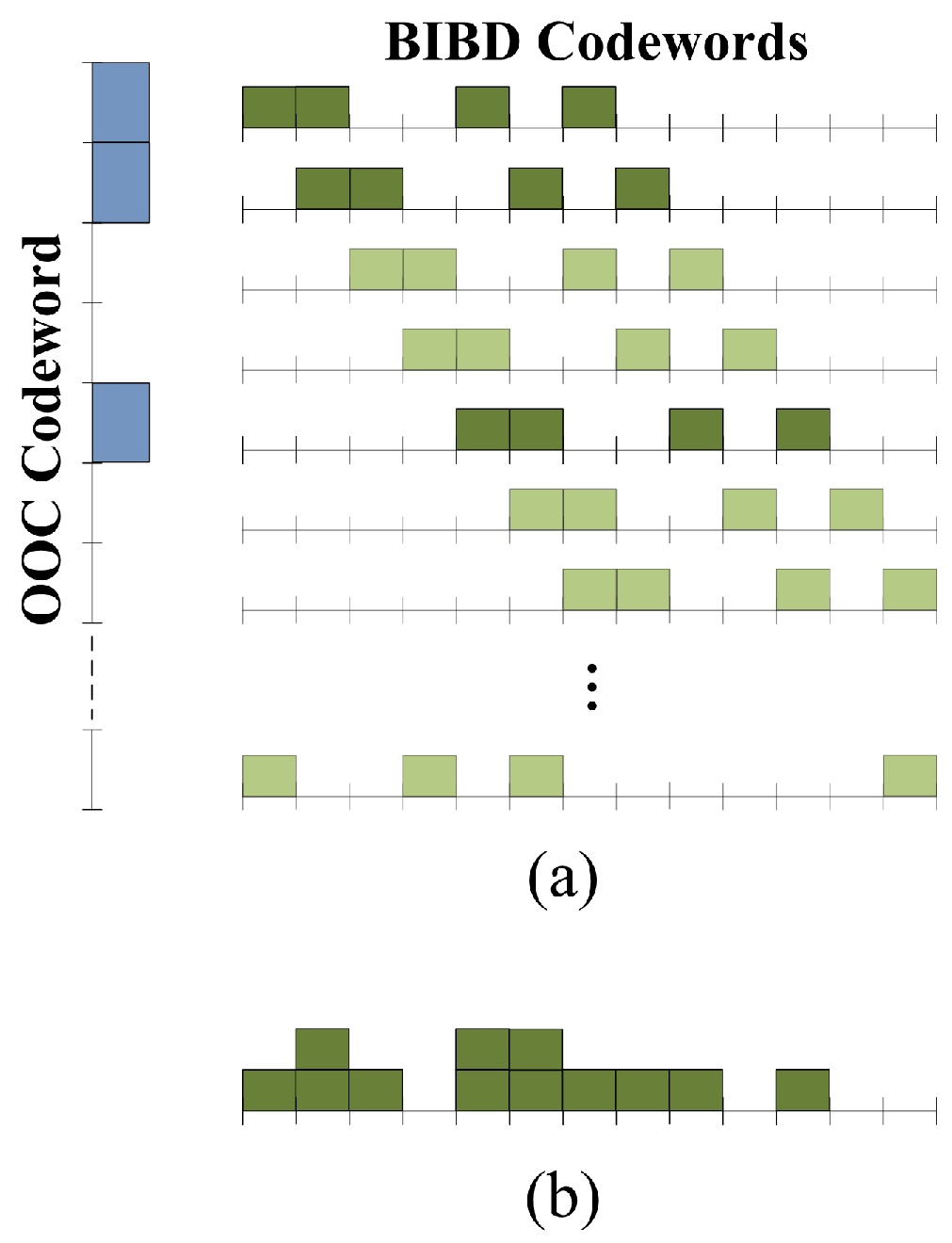}}
    \end{center}
    \vspace*{-0.0 in}
    \caption{(a) Symbol generation for coded-MEPPM using the (1100100000000) OOC codeword and a (13,4,1)-BIBD, and (b) the resulting symbol.}
    \label{C-MEPPM Symbol}
    \end{figure}
    \begin{figure} [!t]
    \vspace*{-0.0 in}
    \begin{center}
    {\includegraphics[width=3.0in]{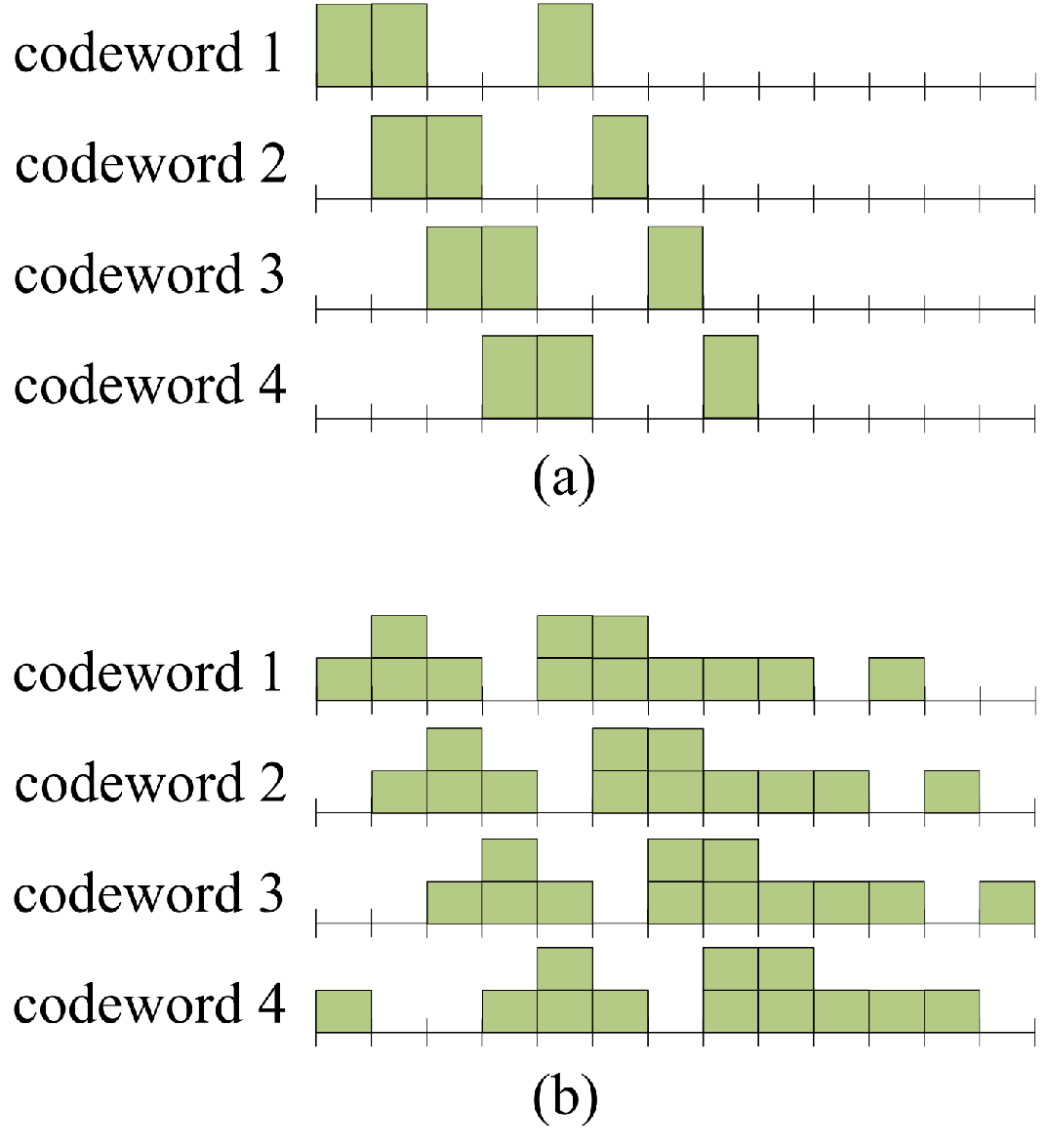}}
    \end{center}
    \vspace*{-0.0 in}
    \caption{First four symbols of one user for (a) CCM-OCDMA using the (1100100000000) OOC codeword, and for (b) coded-MEPPM using the same OOC codeword and the (13,4,1)-BIBD shown in Fig.~\ref{C-MEPPM Symbol}}
    \label{C-MEPPM vs CCM-OCDMA Symbols}
    \end{figure}

\subsection{Networking Using Coded-MEPPM}
OOCs are used in OCDMA networks to provide simultaneous access for users by assigning signature pattern to each user \cite{OOC-89}. In our proposed technique that we call coded-MEPPM (C-MEPPM), we combine OOCs with MEPPM to provide multi-access and high data-rate for each user. Unlike traditional OCDMA systems where OOCs are applied in the time domain, in our approach the OOC codewords are implemented in the code domain, and are applied on the codewords of a BIBD code in code-space.

Let $\mathbf{d}_n=[d_{n1},d_{n2},\dots,d_{nL}]$, $n=1,2,\dots,N$, $d_{n\ell} \in \{0,1\}$, be the $n$th codeword of an OOC with length $L$, weight $w$ and cross-correlation $\alpha$. By assigning the $n$th OOC codeword to user $n$ and assuming $N$ active users, the $m$th symbol in the $Q$-ary constellation is given by
    \begin{align}\label{coded-MEPPM Symbols}
        \mathbf{u}_{m,n} = \frac{1}{Nw} \sum_{\ell=1}^L d_{n \ell} \, \mathbf{c}{^{(m)}_{\ell}}.
    \end{align}
%
In this manner, the symbols of user $n$ are cyclic shifts. The factor $\frac{1}{Nw}$ in (\ref{coded-MEPPM Symbols}) guarantees a PAPR of $Q/K$ for LED arrays. From (\ref{coded-MEPPM Symbols}), the length of the OOC should be no longer than the number of BIBD codewords, i.e., $L \leq Q$. In this work, for a fixed $Q$ we use an OOC with $L=Q$, since the larger the OOC-length, the higher performance it can achieve. Fig.~\ref{C-MEPPM Transmitter}(b) shows the transmitter structure using shift-registers for coded-MEPPM. To illustrate the concept, Fig.~\ref{C-MEPPM Symbol} shows the multilevel symbol generated using a (13,3,1) OOC codeword and a (13,4,1) BIBD code. The first four symbols of CCM-OCDMA using the OOC codeword and coded-MEPPM described in Fig.~\ref{C-MEPPM Symbol} are shown in Fig.~\ref{C-MEPPM vs CCM-OCDMA Symbols}.  Note that CCP-OCDMA is two-level while C-MEPPM is multilevel.

Using the C-MEPPM technique, the dimming level can be controlled by changing the BIBD code as discussed in \cite{VLC-JLT-I-13}, and the maximum number of users can be increased by switching the OOC code or by using a time-sharing technique. Therefore, a significantly smaller code database is needed compared to the case when only OOCs are used for networking in VLC systems.

\subsubsection{Decoders for Multiuser C-MEPPM System}

    \begin{figure} [!t]
    \vspace*{-0.0 in}
    \begin{center}
    {\includegraphics[width=3.3in]{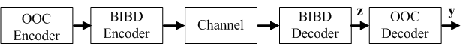}}
    \end{center}
    \vspace*{-0.1 in}
    \caption{Schematic view of a code-MEPPM system using OOC.}
    \label{OOC-BIBD}
    \end{figure}
    \begin{figure} [!t]
    \begin{center}
    {\includegraphics[width=3.3in]{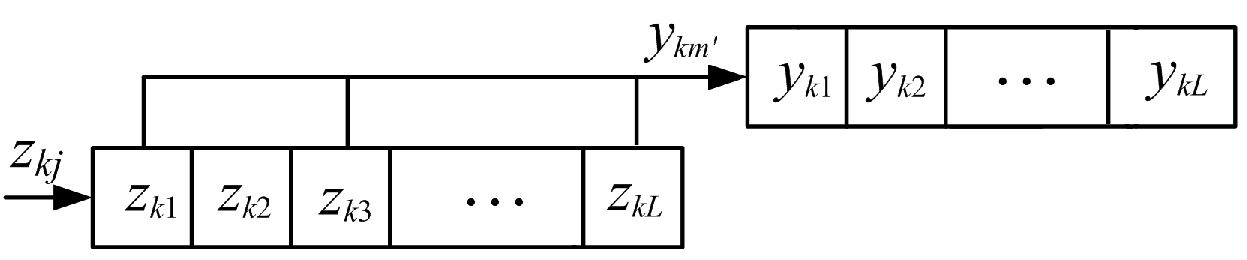}}
    \end{center}
    \vspace*{-0.0 in}
    \caption{OOC decoder using a shift register.}
    \label{OOC Decoder}
    \vspace*{-0.1 in}
    \end{figure}

In a VLC system, due to the short range of the channel, the received power is usually high, and therefore, the system is limited by shot-noise. In \cite{VLC-JLT-I-13} we show that in single-user systems the optimum decoder for EPPM in the shot-noise limited regime is the correlation receiver, and hence, C-MEPPM symbols can be suboptimally decoded using two successive shift registers as shown in Fig.~\ref{OOC-BIBD}. Both consider a single-user environment ignoring the multiple-access interference (MAI). To implement this sub-optimum correlation detector, at the receiver side a BIBD-correlator (Fig.~\ref{EPPM Decoder}) is followed by a OOC-correlator (Fig.~\ref{OOC Decoder}) to decode the received signal. The output vector of the BIBD-correlation decoder, $(y_{k1},y_{k2},\dots,y_{kL})$, resembles the received signal of an CCM-OCDMA system using OOC. The decision is made in favor of the symbol that has the largest corresponding output of the OOC-decoder, i.e.,
    \begin{align}\label{ML Decoder for C-MEPPM}
        \hat{m}_k &= \arg \max_{1 \le m \le L} y_{km}.
    \end{align}

The structure of the optimum multiuser detector (MUD) is highly complex for shot-noise limited multiple access systems \cite{OCDMA_MUD_MBP-95}, \cite{OCDMA_MUD_Poor-95}. Therefore, in this section we derive the optimum single-user detector (SUD) for coded-MEPPM when the system performance is limited by shot-noise. We consider the mean of the interference as a constant background light, since considering it as a modulated signal would require a multiuser detector. The optimum SUD for user $n$ assuming a background photon-count of $\Lambda'_b=\Lambda_b+\frac{N-1}{NQ}\Lambda_0 K$ in each time-slot is
    \begin{align}\label{ML Decoder for C-MEPPM}
        \hat{m}_k &= \arg \max_{1 \le m \le Q} \Bigg( \sum_{j=1}^Q {r_{kj}} \log \Big(\Lambda_0 \mathbf{u}_{m,n}(j) + \Lambda'_b\Big) \nonumber\\
                &\hspace{2.5 cm}- \sum_{j=1}^Q \Big(\Lambda_0 \mathbf{u}_{m,n}(j)+\Lambda'_b \Big)  \Bigg).
    \end{align}
where $\Lambda_0$ is the total received photoelectron-count from the lighting source in one time-chip, $\Lambda_b$ is the received photoelectron-count due to the background radiation in one time-chip, and $\mathbf{u}_{m,n}(j)$ is component $j$ of vector $\mathbf{u}_{m,n}$. $\Lambda_0$ can be written as $\Lambda_0=\eta \frac{P_0}{h \nu} \frac{T_s}{Q}$, where $P_0$ is the peak received power, $h$ is Planck's constant, $\nu$ is the central optical frequency, and $\eta$ is the efficiency of the photodetector.
Considering that both OOC and BIBD are fixed-weight codes, the optimum decoder can be simplified to
    \begin{align}\label{}
        \hat{m}_k &= \arg \max_{1 \le m \le Q} \Bigg( \sum_{j=1}^Q {r_{kj}} \log \Big(N \mathbf{u}_{m,n}(j) \nonumber\\
        &\hspace{3.2 cm}+ N\frac{\Lambda_b}{\Lambda_0} + (N-1)\frac{K}{Q} \Big).
    \end{align}
Defining $\mathbf{v}_n := (v_{1,n},v_{2,n},\dots,v_{Q,n})$ as
    \begin{align}\label{}
        v_{j,n} := \log \Big(\frac{1}{w}\sum\limits_{\ell=1}^{L} d_{n\ell} c_{\ell j} + N\frac{\Lambda_b}{\Lambda_0} + (N-1)\frac{K}{Q}\Big),
    \end{align}
for $j=1,2,\dots,Q$ and $n=1,2,\dots,N$, the optimum detector for user $n$ can be written as
    \begin{align}\label{Optimum Detector Final}
        \hat{m}_k &= \arg \max_{1 \le m \le Q} \big\langle \mathbf{r}_k, \mathbf{v}{^{(m)}_n} \big\rangle,
    \end{align}
and therefore, the optimum SUD is also a correlation detector that can be implemented using a shift-register.

\subsubsection{Error Probability Analysis}

As we shown below, the symbol error rate of the sub-optimum correlation receiver is close to the error rate of the optimum SUD, and acts as a bound on its performance. Therefore, and for mathematical simplicity, we only analyze the error probability of the sub-optimum correlation receiver (Fig.\ref{OOC-BIBD}) in this section.

We derive an approximate mathematical expression for the error-probability of the sub-optimum correlation receiver. The performance of the receiver can be evaluated in three cases: (1) Shot noise limited case, (2) weak MAI scenario, and (3) strong MAI. For the first case we consider a high SNR regime and calculate an upper bound on the symbol error probability. For the second and third cases, MAI is the main limiting factor, and the effect of the shot-noise is less important. For case (3), the performance of C-MEPPM is similar to CCM-OCDMA, and the analytical error probability presented in \cite{OCDMA-M-ary-06} can be used to calculate the BER. For $\alpha = 1$, which corresponds to the weak interference case (case (2)), the approximation given in \cite{OCDMA-M-ary-06} is not accurate. In order to get a more accurate analysis for this scenario, we model the interference as a Bernoulli trial, which has binomial distribution.

For the first case, without loss of generality we assume that the desired user is user 1 and it transmits $\mathbf{s}_{k,1}$ at symbol-time $k$. The overall system using OOC and MEPPM encoders and decoders is depicted in Fig.~\ref{OOC-BIBD}.
Then the symbol error probability, $P_s$, is bounded by
    \begin{align}\label{Union Bound for SER in C-MEPPM}
        P_s \leq \sum_{m_1=1}^{Q} \text{Pr}&(\mathbf{s}_{k,1}=\mathbf{u}_{m_1,1}) \times\nonumber\\
         &\sum\limits_{m' \neq m_1} \text{Pr}(y_{km'}>y_{km_1}|\mathbf{s}_{k,1}=\mathbf{u}_{m_1,1}),
    \end{align}
which is derived using the union bound. For equiprobable transmission,
%
    \begin{align}\label{Pr y_i > y_m1}
        \text{Pr}&(y_{km'}>y_{km_1}  \big| \mathbf{s}_{k,1}=\mathbf{u}_{m_{1},1}) = \frac{1}{L^{N-1}} \sum_{m_2=1}^L \dots \sum_{m_N=1}^L \nonumber\\ & \hspace{1.5cm} \text{Pr}\Big(y_{km'}>y_{km_1}\big| \mathbf{s}_{k,n}=\mathbf{u}_{m_{n},n}, n=1,\dots,N \Big) .
    \end{align}
Here,
    \begin{align}\label{}
        y_{km_1}-y_{km'}=\sum\limits_{j=1}^{Q} r_{kj} \big(\sum\limits_{\ell=1}^{L}d_{1\ell}(c_{(m_1\oplus \ell)j}-c_{(m'\oplus \ell)j}) \big),
    \end{align}
where $\oplus$ is the sum modulo $Q$.  Since $r_{kj}$'s are Poisson distributed random variables, $ \big(y_{km_1}-y_{km'}\big)$ is the sum of weighted Poisson random variables. According to \cite{Weighted-Sum-P-rv}, its probability distribution function (pdf) can be approximated by a Gaussian distribution. Thus, in order to calculate the error probability in (\ref{Pr y_i > y_m1}), we need to calculate the conditional mean and variance of $(y_{km_1}-y_{km'})$.

Given that $\mathbf{s}_{k,n}$ is transmitted for user $n$, $n\in \{1,2,\dots,N\}$, in symbol-time $k$, the mean value of the output of the OOC-correlator that matches the OOC-codeword $\mathbf{d}_1$ in symbol-time $k$, $y_{km'}$, $m'=1,2,\dots,L$, is \cite{VLC-Networking13}
    \begin{align}\label{}
        E&\Big[y_{km'} \Big| \mathbf{s}_{k,n}=\mathbf{u}_{m_{n},n}, n=1,2,\dots,N \Big] =\nonumber\\
        & \Big(\frac{\Lambda_0 K}{Nw}\Big) \langle \mathbf{d}{_{1}^{(m_1)}}, \mathbf{d}{_{1}^{(m')}} \rangle + \underbrace{ \Big(\frac{\Lambda_0 K}{Nw}\Big) \sum_{\substack{{n=2}}}^N \langle \mathbf{d}{_{n}^{(m_{n})}}, \mathbf{d}{_{1}^{(m')}} \rangle}_{\text{MAI}} ,
    \end{align}
and thus,
    \begin{align}\label{}
        E&\Big[y_{km_1}-y_{km'} \Big| \mathbf{s}_{k,n}=\mathbf{u}_{m_{n},n}, n=1,2,\dots,N \Big] =\nonumber\\
        & \Big(\frac{\Lambda_0 K}{Nw}\Big) \bigg(w-\Big\langle \mathbf{d}{_{1}^{(m_1)}}, \mathbf{d}{_{1}^{(m')}} \Big\rangle \bigg) +\nonumber\\& \hspace{1cm} \Big(\frac{\Lambda_0 K}{Nw}\Big) \sum_{\substack{{n=2}}}^N \Big\langle \mathbf{d}{_{n}^{(m_{n})}}, \big( \mathbf{d}{_{1}^{(m_1)}} - \mathbf{d}{_{1}^{(m')}}\big) \Big\rangle ,
    \end{align}
%

In high SNR regimes, an approximate value can be calculated for $\text{Pr}(y_{km'}>y_{km_1}| \mathbf{s}_{k,1}=\mathbf{u}_{m_{1},1})$ by only considering the largest terms in (\ref{Pr y_i > y_m1}), which correspond to the smallest mean and maximum variance of $(y_{km_1}-y_{km'})$ given $\mathbf{s}_{k,1}=\mathbf{u}_{m_{1},1}$. $E[y_{km_1}-y_{km'}| \mathbf{s}_{k,1}=\mathbf{u}_{m_{1},1}]$ takes its minimum value when $\langle \mathbf{d}{_{n}^{(m_n)}},\mathbf{d}{_{1}^{(m_1)}} \rangle=0$ and $\langle \mathbf{d}{_{n}^{(m_n)}},\mathbf{d}{_{1}^{(m')}} \rangle=\alpha$, $n=2,\dots,N$.
For each $n$ at most $ w^2/\alpha $ $m'$'s exist that satisfy the second condition. Defining $\alpha_{ij}:= \langle \mathbf{d}{_{1}^{(i)}},\mathbf{d}{_{1}^{(j)}} \rangle$, in the worst case we have
    \begin{align}\label{}
        E[y_{km_1}-y_{km'}|& \mathbf{s}_{k,1}=\mathbf{u}_{m_{1},1}] \ge \nonumber\\ & \Big(\frac{\Lambda_0 K}{Nw}\Big)(w-\alpha_{m_1m'}+\alpha-\alpha N),
    \end{align}
and this is minimized when $\alpha_{m_1m'}=\alpha$. We thus define
    \begin{align}\label{}
        \mu &:= \min_{m_1 \neq m'} E[y_{km_1}-y_{km'}|  \mathbf{s}_{k,1}=\mathbf{u}_{m_{1},1}] \nonumber\\  &= \Big(\frac{\Lambda_0 K}{Nw}\Big)(w-\alpha N).
    \end{align}

The variance of $\big( y_{km_1}-y_{km'} \big)$ can be calculated as
    \begin{align}\label{}
        \text{Var}&\big[y_{km'}-y_{km_1}| \mathbf{s}_{k,1}=\mathbf{u}_{m_{1},1}\big] =\nonumber\\ &\sum\limits_{j=1}^{Q} \text{Var}[r_{kj}] \Big(\sum\limits_{\ell=1}^{L}d_{1\ell}(c_{(m'\oplus \ell)j}-c_{(m_1\oplus \ell)j}) \Big)^2\nonumber\\
        &= \sum\limits_{j=1}^{Q} \frac{\Lambda_0}{Nw} \sum_{n=1}^N \sum_{\ell'=1}^L d_{n \ell'} c_{(\ell' \oplus m_n)j} \nonumber\\ &\hspace{2cm} \times \Big(\sum\limits_{\ell=1}^{L}d_{1\ell}(c_{(m'\oplus \ell)j}-c_{(m_1\oplus \ell)j}) \Big)^2
    \end{align}
The maximum $\text{Var}\big[y_{km'}-y_{km_1}| \mathbf{s}_{k,1}=\mathbf{u}_{m_{1},1}\big]$ can be written as follows
    \begin{align}\label{}
        \sigma^2 &:= \max_{m_1 \neq m'} \text{Var}\big[y_{km'}-y_{km_1}| \mathbf{s}_{k,1}=\mathbf{u}_{m_{1},1}\big] \nonumber\\ &= \Lambda_0 2 \lambda w(w-1) + \mu.
    \end{align}

Assuming a high SNR regime, we get
    \begin{align}\label{Pe for small N}
        P_s \leq \frac{w^2}{2\alpha} \text{erfc} \Big( \frac{\mu}{\sqrt{2}\sigma} \Big).
    \end{align}
where $\text{erfc}(\cdot)$ is the complementary Gaussian error function. This approximation is not valid when $\alpha N<w$ since in that case $E[ y_{km_1}-y_{km'}]$ can be smaller than zero, which means $\text{Pr}(y_{km'}>y_{km_1}| \mathbf{s}_{k,1}=\mathbf{u}_{m_{1},1})$ approaches 1. This corresponds to the second and third cases, where the MAI is the limiting factor.

As mentioned above, for the second scenario we use a Bernoulli model to approximate the symbol error probability. Denote the probability that a pulse from an interfering user overlaps with one of the pulses of symbol $m'$ and not with any pulses from symbol $m_1$ as $p:=w(w-\alpha)/L$. Then, assuming no shot-noise, $\text{Pr}(y_{km'}>y_{km_1}| \mathbf{s}_{k,1}=\mathbf{u}_{m_{1},1})$ can be written as

    \begin{align}\label{}
        \text{Pr}(y_{km'}&>y_{km_1}| \mathbf{s}_{k,1}=\mathbf{u}_{m_{1},1}) = \nonumber\\ &\sum_{j=0}^{ N-w} \binom{ N-1}{j} p^j (1-p)^{N - 1 - j} \nonumber\\ & \;\;\;\times\sum_{i=w-1+j}^{N-1} \binom{N-1}{i} p^i (1-p)^{ N - 1 - i} ,
    \end{align}
which, for $p \ll 1 $, becomes
    \begin{align}\label{}
        \text{Pr}(y_{km'}>y_{km_1}|& \mathbf{s}_{k,1}=\mathbf{u}_{m_{1},1}) \approx \nonumber\\& \binom{ N-1}{w-1} p^{(w-1)} (1-p)^{( N - w)} .
    \end{align}
As mentioned before, at most $w^2$ symbols of user 1 can have a cross-correlation of $1$ with symbol $m_1$. Thus,
    \begin{align}\label{Pe for large N}
        P_s \approx w^2 \binom{N-1}{w-1} p^{(w-1)} (1-p)^{(N - w)} .
    \end{align}
%

\begin{figure} [!t]
    \vspace*{-0.0 in}
    \begin{center}
    {\includegraphics[width=3.4in]{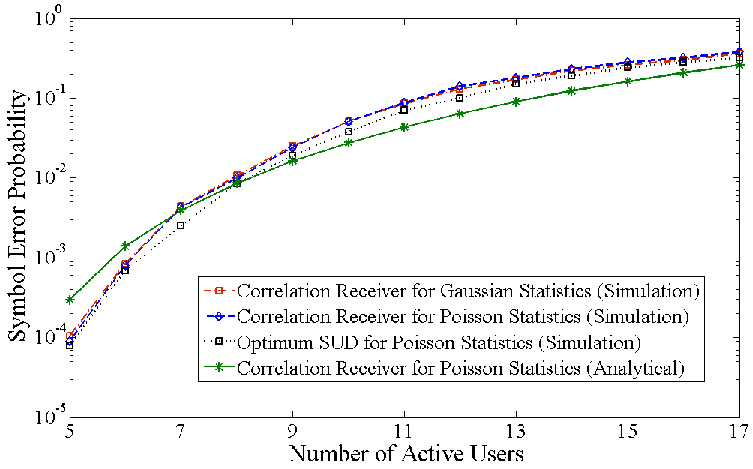}}
    \end{center}
    \vspace*{-0.1 in}
    \caption{BER versus the number of active interfering users for optimum detector from (\ref{Optimum Detector Final}) for Poisson statistics, and correlation receiver for both Poisson and Gaussian statistics. (\ref{Pe for large N}) is used for the analytical results.}
    \label{Different Detectors}
    \end{figure}

Simulation and analytic results for the two detectors are given in Fig.~\ref{Different Detectors} for a coded-MEPPM system using a (341,5,1)-OOC code \cite{OOC-89} and a (341,85,21)-BIBD \cite{BIBD-website}, which can support up to 17 simultaneous users. The optimum SUD derived in (\ref{Optimum Detector Final}) is simulated assuming Poisson statistics, and the correlation detector in Figure~\ref{C-MEPPM Transmitter} is simulated using both Gaussian and Poisson noise. The analytical BER expression from (\ref{Pe for large N}) is also plotted. In these results, both the background light power and the peak received signal power are assumed to be $0.1 \, \mu W$, and the data is assumed to be transmitted using ideal LEDs with a central wavelength of $650$ nm at a bit-rate of $200$ Mb/s through an ideal channel. According to these results, the BER of the optimum SUD is slightly lower than that of the correlation receiver. The Gaussian approximation for the statistics of the received photoelectron count is accurate, as the simulated BER are shown to be close. As can be seen, the analytical result calculated in (\ref{Pe for large N}) provides a good approximation to the BER of the C-MEPPM system.

\subsection{Networking Using Divided-MEPPM}
In the second proposed technique, which we call divided-MEPPM (D-MEPPM), the generating BIBD code is divided into several smaller codes, and a different set of codewords is assigned to each user. Then, as in the MEPPM scheme, each user uses its codeword set to generate multilevel symbols.

Let $q_n$ be the number of BIBD codewords that are assigned to user $n$, such that $q_1+q_2+\dots+q_N=Q$, and let $\mathcal{C}_n$, $ |\mathcal{C}_n |=q_n$, be the set of codewords that are assigned to user $n$, such that $\mathcal{C}_n \cap \mathcal{C}_m = \emptyset$ for any $n \neq m$, and $\mathcal{C}_1 \cup \mathcal{C}_2 \cup \dots \cup \mathcal{C}_N = \{\mathbf{c}_1,\mathbf{c}_2,\dots,\mathbf{c}_Q\} $. This can be considered as a kind of CDMA, where distinct codeword sets with cross-correlation $\lambda$ are assigned to users. Using this definition, user $n$ can utilize an $\ell_n$-branch MEPPM, $1 \leq \ell_n < q_n$ , for $M$-ary transmission using $\mathcal{C}_n$. It can use either MEPPM type-I or type-II, yielding a constellation of size ${q_n}\choose{\ell_n}$ for type-I and ${q_n+\ell_n}\choose{\ell_n}$ for type-II MEPPM \cite{Multilevel-EPPM12}. Symbol $m$ of user $n$ can be expressed as
    \begin{align}\label{}
        \mathbf{u}_{m,n} = \frac{1}{Q} \sum_{\ell \in \mathcal{C}_{m,n}} \mathbf{c}_{\ell},
    \end{align}
where $\mathcal{C}_{m,n}$ is the set of $\ell_n$ codewords assigned to symbol $m$ of user $n$. For type-I MEPPM, all $\ell_n$ elements of $\mathcal{C}_{m,n}$ are distinct, and for type-II, a codeword can be repeated more than once in $\mathcal{C}_{m,n}$. Each user can generate these symbols using the transmitter shown in Fig.~\ref{C-MEPPM Transmitter}-(a). The correlation receiver shown in Fig.~\ref{EPPM Decoder} is used to decode the received signal, and then the detection techniques described in \cite{Multilevel-EPPM12} are used to estimate the received symbol.

An advantage of this networking technique over the one presented in Section III-A is its potential to provide different data-rates for different users, which can be done by assigning unequal size subsets to users. Thus, a larger number of BIBD codewords is provided to users requiring a higher bit-rate. Table~\ref{D-MEPPM List} gives examples of how a system using two possible BIBD codes, (63, 31, 15) and (57, 8, 1) can provide PAPR levels of 2 and 7.125, respectively, while simultaneously assigning each of 5 users in the network various data-rates. Data-rates of users can be changed by assigning code sets of different sizes.
    \begin{table}[!t]
    \caption{Allocations of Codewords to Provide Different Data-rates and PAPRs in a VLC Network}\label{D-MEPPM List}
        \vspace*{-0.2 in}
        \begin{center}
        (63, 31, 15) BIBD\\
            \begin{tabular}{|c||ccccc|}\hline
            User            &   1   &   2     &   3    &   4   &   5   \\
            \hline
            \hline
            Number of Codes &   13   &   13     &   13    &   12   &   12   \\
            Data Rate (Mb/s)&   26   &   26     &   26    &   24   &   24   \\
            \hline
            Number of Codes &   21  &   21    &   7    &   7   &   7  \\
            Data Rate (Mb/s) &  43.3 &  43.3   &  13.3  &  13.3 & 13.3 \\
            \hline
            Number of Codes &   35  &   9     &   9    &   5   &   5  \\
            Data Rate (Mb/s) &   73.7& 17.6    & 17.6   & 9.2   &   9.2  \\
            \hline
            \end{tabular}
        \end{center}
        \begin{center}
            (57, 8, 1) BIBD\\
            \begin{tabular}{|c||ccccc|}\hline
            User            &   1   &   2     &   3    &   4   &   5   \\
            \hline
            \hline
            Number of Codes &   12   &   12     &   11    &   11   &   11   \\
            Data Rate (Mb/s)&   26.5   &   26.5     &   24.1    &   24.1   &   24.1   \\
            \hline
            Number of Codes &   18  &   18    &   7    &   7   &   7  \\
            Data Rate (Mb/s) &  40.7 &  40.7   &  14.8  &  14.8 & 14.8 \\
            \hline
            Number of Codes &   33  &   8     &   8    &   4   &   4  \\
            Data Rate (Mb/s) &   76.6 & 17.1    & 17.1   & 7.8   &   7.8  \\
            \hline
            \end{tabular}
        \end{center}
        \vspace*{-0.2 in}
    \end{table}

A analytical expression can be calculated for D-MEPPM similar to the one above for C-MEPPM.
Without loss of generality we calculate the symbol error probability for the first user given that $\mathbf{s}_{k,n} = \mathbf{u}_{m_n,n}$ is sent for user $n$. In this case, the variance of $z_{kj}$ is
    \begin{align}\label{}
        Var \big[ z_{kj} \big]   &= \sum_{i=1}^Q c_{ji} Var\big[r_{ki}\big] + \Gamma^2 \sum_{i=1}^Q (1-c_{ji}) Var\big[r_{ki}\big]  \nonumber\\
                        &= \sum_{i=1}^Q c_{ji} E\big[r_{ki}\big] + \Gamma^2 \sum_{i=1}^Q (1-c_{ji}) E\big[r_{ki}\big],
    \end{align}
where
    \begin{align}\label{}
        E\Big[r_{ki}\big| \mathbf{s}_{k,n}=&\mathbf{u}_{m_{n},n}, n=1,2,\dots,N \Big] = \frac{1}{Q} \sum_{n=1}^N \sum_{\ell \in \mathcal{C}_{m_n,n}} c_{\ell i} .
    \end{align}
Since $\mathcal{C}_{m_n,n}$'s are distinct sets we have
    \begin{align}\label{}
        Var&\Big[z_{kj}\big| \mathbf{s}_{k,n}=\mathbf{u}_{m_{n},n}, n=1,2,\dots,N \Big] \nonumber\\ &= Var\Big[z_{kj}\big| \mathbf{s}_{k,1}=\mathbf{u}_{m_{1},1} \Big] \nonumber\\
        &= \frac{\Lambda_0}{Q} \Big[ (\ell_1+\dots+\ell_N)\Big(\frac{\lambda K}{K-\lambda}\Big) +  n_{\mathbf{c}_j} \frac{(K-2\lambda) K}{K-\lambda} \Big].
    \end{align}
where $n_{\mathbf{c}_j}$ is the number of appearance of codeword $\mathbf{c}_j$ in the set $\mathcal{C}_{m_1,1}$.
Here we approximate $Var[ z_{kj} ]$ as
\begin{equation}
        Var[ z_{kj} ] \approx \frac{K \Lambda_0}{Q(K-\lambda)} \Big[ (\ell_1+\dots+\ell_N) \lambda+  K-2\lambda \Big].
\end{equation}
We can then use the following expressions to approximate the BER \cite{Multilevel-EPPM12}:
    \begin{align}\label{}
       P_b \approx  M' \; \text{erfc} \Bigg( \sqrt{ \frac{ (K-\lambda)^2 \Lambda_0/Q}{(\ell_1+\dots\ell_N)\lambda K +  (K-2\lambda) K} } \Bigg) .
    \end{align}
where
\begin{equation}
M'=\frac{\ell_1(q_1-\ell_1)}{8}
\end{equation}
for type-I MEPPM and
 \begin{equation}
 M'=\frac{\ell_1q{_1^2}(q_1-1)^2}{8(q_1+\ell_1)(q_1+\ell_1-1)(q_1+\ell_1-2)}
 \end{equation}  for type-II.

    \begin{figure} [!t]
    \vspace*{-0.0 in}
    \begin{center}
    {\includegraphics[width=3.4in]{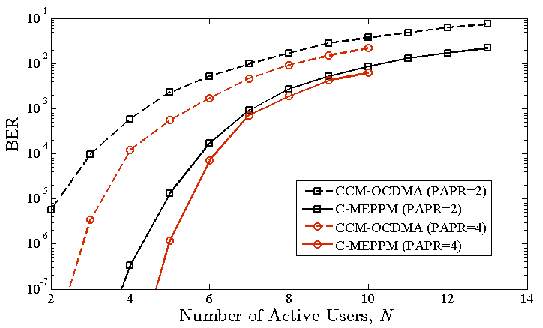}}
    \end{center}
    \vspace*{-0.0 in}
    \caption{BER of C-MEPPM and CCM-OCDMA versus the number of active users for a PAPR of 2 and 4, peak received power $P_0=0.1\; \mu$W, background power $P_b=0.1 \; \mu$W, central wavelength $\lambda=650$, and photo-detector efficiency $\eta=0.8$.}
    \label{C-MEPPM vs CCM-OCDMA.}
    \end{figure}
    \begin{figure} [!t]
    \vspace*{0.0 in}
    \begin{center}
    {\includegraphics[width=3.4in]{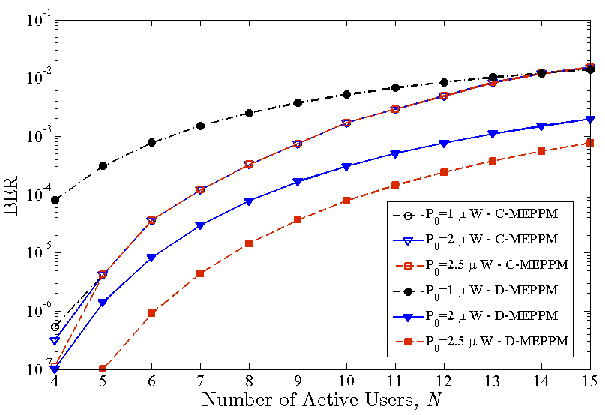}}
    \end{center}
    \vspace*{-0.0 in}
    \caption{Analytical BER versus the number of active interfering users for coded-MEPPM (C-MEPPM) and divided-MEPPM (D-MEPPM) with different peak power levels for $\eta=0.8$, central wavelength $\lambda=650$ and background power $P_b=0.1$ $\mu W$.}
    \label{OWN}
    \end{figure}

\section{Numerical Results}
In this section numerical results using the analytical BER expressions derived above are presented to compare the performance of the proposed networking techniques to each other, and also to that of a CCM-OCDMA system.

Fig.~\ref{C-MEPPM vs CCM-OCDMA.} compares the BER of a CCM-OCDMA system with that of a coded-MEPPM for different numbers of active users and for two PAPR values. For coded-MEPPM, the BIBD code is chosen to satisfy the PAPR condition, and the OOC is chosen to provide the lowest collision probability between users, i.e., the smallest MAI effect. For a PAPR $=4$ and a constellation size of 101, the CCM-OCDMA system uses a (101,25,7)-OOC  and the coded-MEPPM uses a (101,11,2)-OOC and a (101,25,6)-BIBD, for a maximum of 10 users. Since the parameters of the OOC must satisfy $\alpha>w^2/L$ \cite{OOC-89}, for the CCM-OCDMA with $L=101$ and $w=25$, $\alpha=7$ is the smallest cross-correlation that can be chosen. As shown in the figure, coded-MEPPM can achieve a lower BER compared to CCM-OCDMA. For a PAPR $= 2$, the BER of a CCM-OCDMA system using a (83,41,21)-OOC is compared to that of a coded-MEPPM system using a (83,41,20)-BIBD and a (83,15,3)-OOC for different number of active users up to 13 users. The coded-MEPPM is shown to have a better performance compared to CCM-OCDMA.

A comparison between the analytical BER of the coded-MEPPM using a (63,31,15)-BIBD and a (63,7,2)-OOC, and divided-MEPPM using the same BIBD code is shown in Fig.~\ref{OWN}. In this figure the BER is plotted versus the number of users for three different peak received power levels and for a PAPR of 2. The C-MEPPM system can provide simultaneous multiple access for up to 16 users. For divided-MEPPM, distinct sets of 4 BIBD codewords are assigned to users, and therefore, the maximum number of users is again 16. For this technique, each user utilizes 4-level type-II MEPPM and has 70 symbols. According to these results, for weak peak powers the error probability of coded-MEPPM is lower than divided-MEPPM since it has larger distance between its symbols and its performance is only limited by MAI, while the performance of divided-MEPPM is limited by the shot noise.

\section{Conclusion}
Two multiple-access methods for use in VLC systems are introduced and compared. The first method, using both OOC and BIBD codes to encode the user data, is shown to have a large distance between symbols. Unlike CCM-OCDMA systems, the parameters of the OOC that is used in C-MEPPM can be chosen independent from the given PAPR, and hence, C-MEPPM is able to provide a lower BER compared to CCM-OCDMA when MAI is the dominating factor. The second technique only uses BIBD codes to construct MEPPM symbols, and, therefore, can have a large constellation size and consequently a high bit-rate for each user. It is also able to provide flexible and unequal data-rates to users. According to the numerical results, for low SNR cases the coded-MEPPM technique achieves a lower BER compared to divided-MEPPM, while the latter is preferred in high SNR regimes since it has a lower MAI effect.

\section{Acknowledgment}
This research was funded by the National Science Foundation (NSF) under grant number ECCS-0901682.

\bibliographystyle{IEEEtran}
\bibliography{EPPM}

\balance

\end{document}